\documentclass[aps,prb,reprint,unsortedaddress,superscriptaddress,showpacs]{revtex4-1}
\usepackage{amsmath,amssymb,graphicx,color,bm}

\begin{document}

\title{
Criticality of the metal--topological insulator transition driven by disorder
}

\author{Ai Yamakage}
\affiliation{Department of Physics, Tohoku University, Sendai 980-8578, Japan}
\affiliation{Department of Applied Physics, Nagoya University, Nagoya 464-8603, Japan}

\author{Kentaro Nomura}
\affiliation{Department of Physics, Tohoku University, Sendai 980-8578, Japan}
\affiliation{Correlated Electron Research Group (CERG), RIKEN-ASI, Wako 351-0198, Japan}
\affiliation{Institute for Materials Research, Tohoku University, Katahira, Aoba-ku, Sendai 980-8577}

\author{Ken-Ichiro Imura}
\affiliation{Department of Physics, Tohoku University, Sendai 980-8578, Japan}
\affiliation{Department of Quantum Matter, AdSM, Hiroshima University, Higashi-Hiroshima 739-8530, Japan}

\author{Yoshio Kuramoto}
\affiliation{Department of Physics, Tohoku University, Sendai 980-8578, Japan}

\date{\today}

\begin{abstract}
Employing scaling analysis of the localization length, 
we deduce the critical exponent of the metal--topological insulator (TI) transitions induced by disorder.
The obtained exponent $\nu \sim 2.7$ shows no conspicuous deviation from the value established for metal--ordinary insulator transitions in systems of the symplectic class.
We investigate the topological phase diagram upon carrier doping to reveal the nature of the so-called topological Anderson insulator (TAI) region. 
The critical exponent of the metal--TAI transition is also first estimated, shown to be undistinguishable from the above value within the numerical error. 
By symmetry considerations we determine the explicit form of Rashba spin--orbit coupling in systems of $C_{4v}$ point group symmetry.
\end{abstract}

\pacs{72.15.Rn,73.20.Fz,71.70.Ej,73.43.-f}

\maketitle

\section{Introduction}

The concept of the topological insulator (TI) has been widely recognized
in the condensed-matter community 
since the appearance of the 
$\mathbb Z_2$ TI.
\cite{kane05_2, HasanKane, qi11,yan12}
Prototype of the TI in two spatial dimension is a quantum Hall insulator.
In recent years, quantum spin Hall (QSH) insulators have attracted much attention.
The quantum Hall effect requires a finite, rather a strong magnetic field that breaks the time reversal symmetry. 
The QSH effect occurs without a magnetic field, 
induced solely by the spin-orbit coupling (SOC) that preserves time-reversal symmetry. 
\cite{kane05}
The latter TI has been first experimentally
observed in a HgTe/CdTe quantum well,
\cite{konig07}
after a theoretical prediction of Bernevig, Hughes and Zhang (BHZ).
\cite{bernevig06}
Recently, a InAs quantum well has been observed to be a TI.\cite{liu08,knez10,knez11,knez12}

Robustness against disorder
is a defining property of the topological quantum phenomena.
Quantifying the role of disorder has also played
a central role in the conceptual development of the TI.
Random matrices are classified into three categories:
orthogonal, unitary and symplectic, according to their symmetry.
\cite{Dyson}
The distinction between quantum Hall effect and QSH effect might be most accentuated in this context. 
The quantum Hall effect, breaking time-reversal symmetry, belongs to the unitary symmetry class,
while QSH systems are symplectic when rotational symmetry of spin is broken by SOC.
A more exhaustive classification of the disordered systems
\textit{a la} Refs. \onlinecite{AltlandZirnbauer, Zirnbauer}
has led to the periodic table of 
topological insulators and superconductors.
\cite{Schnyder_PRB, Kitaev_AIP, Schnyder_AIP, Ryu_NJP}
QSHs belong to class ``AII"in this classification.

It is naturally an interesting question whether criticality of disorder-induced transition between the metal and the TI suggest the existence of a new universality class reflecting the nontrivial topological nature.
So far researches in this direction have been performed on the
Kane-Mele,
\cite{onoda07} 
and $\mathbb Z_2$ network
\cite{obuse07, obuse08, Koba}
models.
Detailed analyses in Refs. \onlinecite{obuse07, obuse08, Koba}
imply that
the critical behavior of the disordered TI
is undistinguishable from that of the conventional symplectic systems.
\cite{asada02}
These studies, however, does not directly show the relation between their systems and the actual TIs.
Furthermore, recently, localization in weak TIs\cite{mong12,kobayashi12} and  topological crystalline insulators\cite{fu12} have been suggested to show a new criticality of the metal-insulator transition.
Thus more studies on the criticality in TIs are still desired.


Another motivation of this work is to settle down the controversial issues on the so-called topological Anderson insulator (TAI).
Though
the phase diagram of the 
disordered
TIs has been studied in the literature,\cite{onoda07, IKN_EPL, IKN_PRB}
it has been pointed out recently that 
there is a region in the phase diagram dubbed as the TAI region
in which
an ordinary insulator is converted to a TI
solely by disorder.
\cite{li09}
The nature of the TAI has been already much discussed.
\cite{groth09, yamakage11, prodan11_2, prodan11, xu12, Shen2}
For example, ``Is TAI a distinct phase?" \cite{prodan11} has been one of the main questions.
We have partly addressed this question in Ref. \onlinecite{yamakage11} 
from the viewpoint of phase diagram in the ($\Delta$, $W$)-plane,
where $\Delta$ is the mass term determining the topological phase, and $W$ is a measure of the strength of disorder.
Here, on the other hand, we give a closer inspection of the transition by improving the precision of our previous study,
presenting a detailed discussion on the critical exponent characterizing the nature of the TAI.
We restrict our investigations to 2D, but the idea of the TAI has been equally applied to the three dimensional systems.\cite{guo10, goswami11, ryu12}

In this paper, 
we employ a BHZ-like effective Hamiltonian implemented on a square lattice 
reinforced with the Rashba-type $s_z$ non-conserving SOC, which represents a HgTe quantum well.
We deduce the critical exponent $\nu$ of the metal--insulator transitions from scaling--analysis of the localization length of the wave function.
The obtained $\nu \sim 2.7$  coincides with that of the symplectic class.
Our result supports that the critical properties of the actual material can be mapped to the effective network models.
Furthermore, we clarify the phase diagram: 
The TAI is continuously connected to the clean TI in the phase diagram, i.e., the TAI is not a new phase.
Also, we clarify that Rashba SOC and carrier-doping crucially affect on the phase diagram:
metallic phase appears and predominates TI and TAI in a wide phase region.
Moreover, we revisit the nature of TAI.
It has been known that the mass renormalization is a useful description for TAI. 
In the heavily doped case, however, the mass renormalization picture fails to capture the nature of TAI.

The paper is organized as follows.
In Sec. II, a model Hamiltonian of a HgTe quantum well is shown.
The detailed derivation of this Hamiltonian from the symmetry consideration is shown in Appendix \ref{rashba}.
In Sec. III, critical exponent of the metal-insulator transition, which is determined by the finite-size scaling, is discussed. 
Furthermore, effects of carrier-doping and Rashba SOC on the phase diagram is clarified in Sec. IV.
Finally, we summarize our results in Sec. V.




\section{Model}

We start with the following BHZ-like effective Hamiltonian,
\begin{align}
 H(\bm k) = \begin{pmatrix}
  h(\bm k) & \Gamma(\bm k)
  \\
  \Gamma^\dag(\bm k) & h^*(-\bm k)
 \end{pmatrix}.
 \label{hmlt}
\end{align}
expressed in $\bm k$-space.
The basis is as
$(
\left| 1 \uparrow \right\rangle, 
\left| 2 \uparrow \right\rangle, 
\left| 1 \downarrow \right\rangle, 
\left| 2 \downarrow \right\rangle
)^{\rm T}$, 
where 1 and 2 denote the orbitals with even and odd parity under spatial inversion for $\bm k=0$.
The arrows denote the spin up ($\uparrow$) and down ($\downarrow$) of an electron.
The two diagonal blocks correspond to the spin up and down sectors, and
$\Gamma(\bm k)$ represents the spin-flip hopping
induced by the Rashba-type SOC.
The spin up sector of the Hamiltonian takes the Dirac form,
\cite{bernevig06}
\begin{align}
 h(\bm k) = 
 \begin{pmatrix}
   \Delta - B \bm k^2 & -iAk_+
   \\
   iAk_- & -\Delta + B \bm k^2
 \end{pmatrix},
\end{align}
where $k_\pm = k_x \pm i k_y$.
The spin down part is given
by the time-reversal of the spin up part, i.e., $h^*(-\bm k)$.
$\Delta$ denotes the mass term, whose magnitude corresponds to that of the band gap. 
$B$ corresponds to conduction and valence band curvatures.
Here, they are assumed to be
the same.
$A$ is the strength of hybridization between the orbitals.
%
$\Delta /B > 0$  ($\Delta /B < 0$) corresponds
if $A \ne 0$
to the topologically nontrivial (trivial) phase.
Transition between these topologically distinct sectors
occurs at
$\Delta=0$, accompanied by
a band-gap closing at $\bm k = \bm 0$.
%
%
$\Gamma(\bm k)$ represents spin-flip terms given by (See Appendix \ref{rashba})
\begin{align}
 \Gamma(\bm k) =
 \begin{pmatrix}
   i\alpha k_- & 0
   \\
   0 & -i\alpha k_+
 \end{pmatrix},
\end{align}
with the coupling constant $\alpha$ of Rashba SOC.
For a smaller Rashba SOC ($|\alpha| < |A|$), the resulting topological phase diagram is the same as that for $\alpha=0$.
Effect of $\alpha$ on the phase diagram is discussed in Appendix \ref{cleanlimit}.

In the actual calculation, 
the model is implemented on a square lattice.
This corresponds to the substitution,
$k_i \to \sin k_i$ and $k_i^2 \to 2(1-\cos k_i)$.
The phase diagram of this model is different from that of the original effective model since band gap closing (topological phase transition) occurs not only at the Brillouin zone center but also at the zone boundary.
In the clean limit
the system becomes a TI at half filling
if $0<\Delta/B< 4$ or $4 < \Delta/B < 8$.
The band gap closes at $\bm k = (0, \pi)$ and $\bm k = (\pi, 0)$ for $\Delta = 4B$, and at $\bm k = (\pi, \pi)$ for $\Delta = 8B$.

In real space
the tight-binding Hamiltonian as defined above
on the square lattice is expressed as
\begin{align}
\label{tight}
 H = \sum_{\bm r}
 \left[
	c^\dag_{\bm r} \epsilon_{\bm r} c_{\bm r}
	+ \left(
		c^\dag_{\bm r} t_x c_{\bm r + \bm a}
		+
		c^\dag_{\bm r } t_y c_{\bm r + \bm b}
		+
		\mathrm{h.c.}
	\right)
 \right],
\end{align}
where $c_{\bm r} = (c_{1 \uparrow}, c_{2 \uparrow}, c_{1 \downarrow}, c_{2 \downarrow})^{\rm T}$ denotes the annihilation operator of an electron at $\bm r$, and $\bm a = (1,0)$, $\bm b = (0,1)$ are the primitive translational vector. 
Here the lattice constant is set to be unity.
$t_x$ and $t_y$ are hopping amplitudes
in the $x$ and $y$ directions, and are given as
\begin{align}
 t_x &=
 B \sigma_z s_0 - i \frac{A}{2} \sigma_y s_0 + i \frac{\alpha}{2} \sigma_z s_y,
 \\
 t_y &=
 B \sigma_z s_0 - i \frac{A}{2} \sigma_x s_z - i \frac{\alpha}{2} \sigma_0 s_x.
\end{align}
$\sigma_i$ and $s_i$ with $i=x,y,z$ are Pauli matrices and $\sigma_0$ and $s_0$ are identity matrices in the orbital and spin spaces.
Randomness by disorder is incorporated in the on-site potential $\epsilon_{\bm r}$ as
\begin{align}
\epsilon_{\bm r} = 
\begin{pmatrix}
  \Delta - 4B + W_{\bm r}^{+} & 0
  \\
  0 & -\Delta + 4B + W_{\bm r}^{-}
\end{pmatrix},
\label{dis}
\end{align}
where
$W_{\bm r}^{\pm}$ is a probability variable obeying the uniform distribution which takes a value within $[-W/2, W/2]$.
$W$ corresponds to the strength of disorder.
In the numerical computation, the parameters are set as $A=B=1$, for simplicity.

Let us comment on the
symmetries of the system.
In the absence of Rashba SOC, 
the total Hamiltonian is 
divided into two decoupled
spin up and down sectors. As a result, the system shares the same unitary class (A) with
those exhibiting the quantum Hall effect.
In the presence of Rashba SOC, on contrary,
the system has time-reversal symmetry as $\Theta H \Theta^{-1} = H$ with $\Theta = -i s_y \mathcal K$, $\Theta^2 = -1$, 
and $\mathcal K$ being complex conjugation.
Then, the system belongs to the symplectic class (AII).
At $W=0$,
our model has also a chiral symmetry, {\it i.e.},
$\Gamma H \Gamma^\dag = -H$ with $\Gamma = \sigma_x s_y$.
In this case, 
combining $\Theta$ and $\Gamma$,
one can construct a particle-hole operator such that
$\mathcal C H \mathcal C^{-1} = -H$ with $\mathcal C = \Gamma \Theta = \sigma_x \mathcal K$ and $\mathcal C^2=1$.
The chiral (or particle-hole) symmetry imposes
the energy spectrum to be symmetric with respect to $E=0$.
On turning on the disorder ($W \ne 0$)
that breaks microscopically the chiral symmetry,
the chiral symmetry is generally broken,
and the corresponding energy band becomes
asymmetric with respect to $E=0$.
In this case,
the system turns to fall on the symplectic class (AII).
Only at $W=0$, {\it i.e.},
in the presence of both the time-reversal and chiral symmetries,
the system belongs to the class DIII.
Note that breaking of chiral symmetry may affect the behavior of 
gapless/gapped edge states.\cite{mao11}

\section{Criticality}
Employing scaling analysis of the localization length, 
we calculate the critical exponent of 
metal--TI
transitions driven by disorder.
After a brief sketch of the method,
we show extensive data on the critical exponents, 
and compare our results with those for other related models 
in the light of universality class arguments.
These should be regarded as central results of the paper.

\subsection{Localization length and finite-size scaling}

We briefly review our method to determine a critical point and the corresponding exponent from finite-size scaling.\cite{abrahams79,mackinnon81,mackinnon83,kramer93,markos06}
In disordered quasi-one dimensional systems, 
wave functions $\psi_i(x)$'s decay as $\psi_i(x) \sim e^{\pm x/\lambda_i}$, where $\lambda_i$ ($\lambda_1 \leq \cdots \leq \lambda_L$) is the localization length of the system.
The largest localization length $\lambda_L$ diverges at the critical point of metal-insulator transition as $\lambda_L \sim |q - q_{\rm c}|^{-\nu}$, where $q$ is a parameter representing disorder strength $W$ or energy $E$, and $q_{\rm c}$ is the critical value.
The largest localization length divided by the system width $\lambda_L/L$, where $L$ is the width of the system,  becomes scale-free near the critical point.
Therefore, $\lambda_L/L$ is expanded  as
\begin{align}
 \frac{\lambda}{L}
 = \Lambda_0 + \sum_{n=1}^{N} a_n (q-q_{\rm c})^n L^{n/\nu}
 + \sum_{n=0}^{M} b_n (q-q_{\rm c})^n L^{n/\nu+y},
 \label{fss}
\end{align}
The third term in the above expression is an irrelevant length scale collection with a negative irrelevant exponent $y$.
In $L \to \infty$, the single-parameter scaling recovers:
\begin{align}
 \frac{\lambda}{L} \to \Lambda' = \Lambda_0 + \sum_{n=1}^N a_n(q-q_{\rm c})^n L^{n/\nu}.
\end{align}
The present system has $2gL$ localization lengths, where $g=4$ is the number of internal degrees of freedom at each site.
These are calculated by the transfer-matrix method (See Appendix \ref{method}).
The numerical data of the largest localization length is fitted by Eq. (\ref{fss}).
As a result, the critical value $q_{\rm c}$ and the critical exponent $\nu$ are obtained.
In the actual calculation, we fix the fitting parameters to $N=4$ and $M=2$.
Additionally, some remarks on the size-scaling is discussed in Appendix \ref{remarks}.

\subsection{Critical exponent}

\begin{figure*}
\centering
\includegraphics{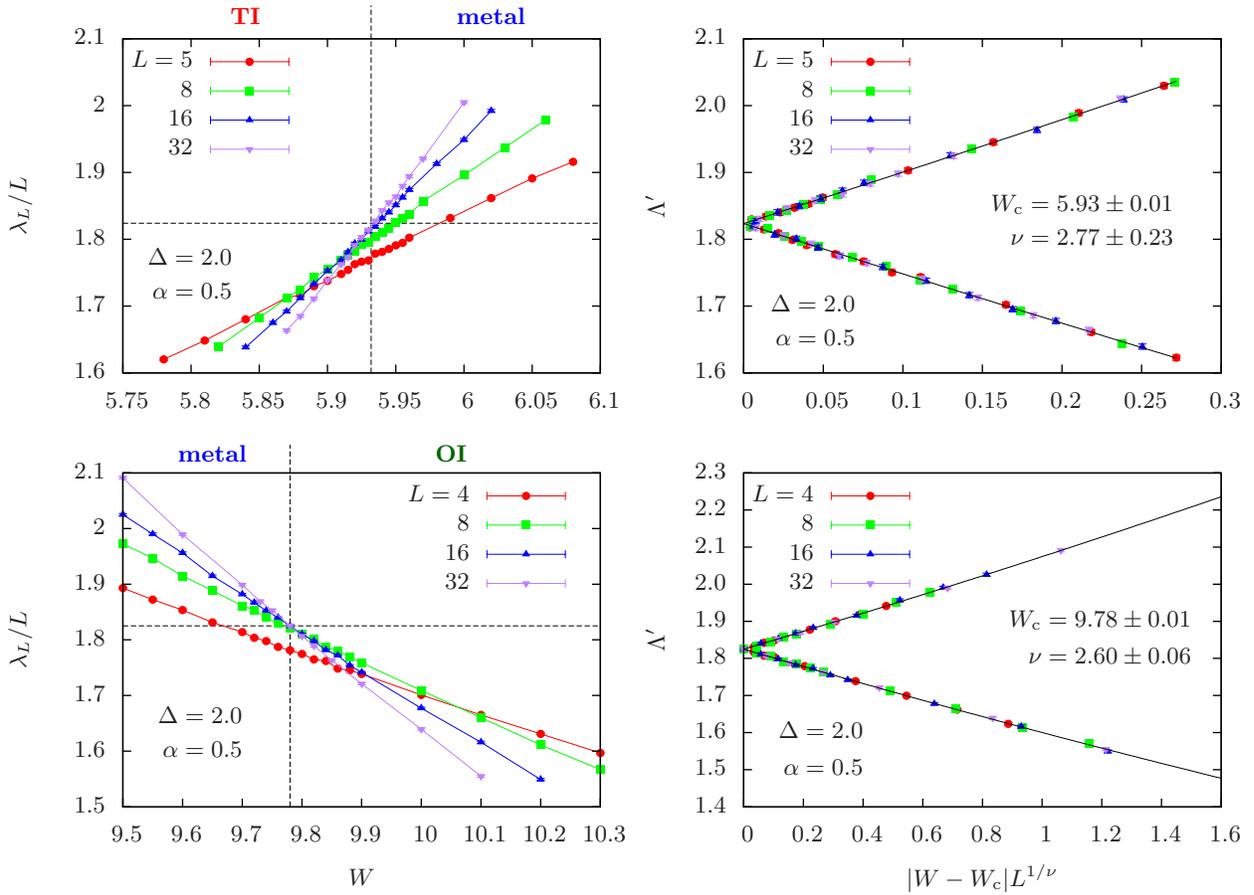}
\caption{Localization length divided by the system size ($\lambda_L/L$) (left) and the single-parameter-scaling part $\Lambda'$ (right) for $\Delta=2.0$ and $\alpha=0.5$.
The upper (lower) panels show the localization length in the vicinity of the transition point from a topological (ordinary) insulator to a metal.
The horizontal axis corresponds to $\left| W-W_{\rm c} \right| L^{1/\nu} \propto (L/\xi)^{1/\nu}$ with a correlation length $\xi$.
The dashed lines in the left panels denote the critical point determined from the the finite size scaling shown in the right panels.
}
\label{critical_fig}
\end{figure*}

\begin{figure*}
\centering
\includegraphics{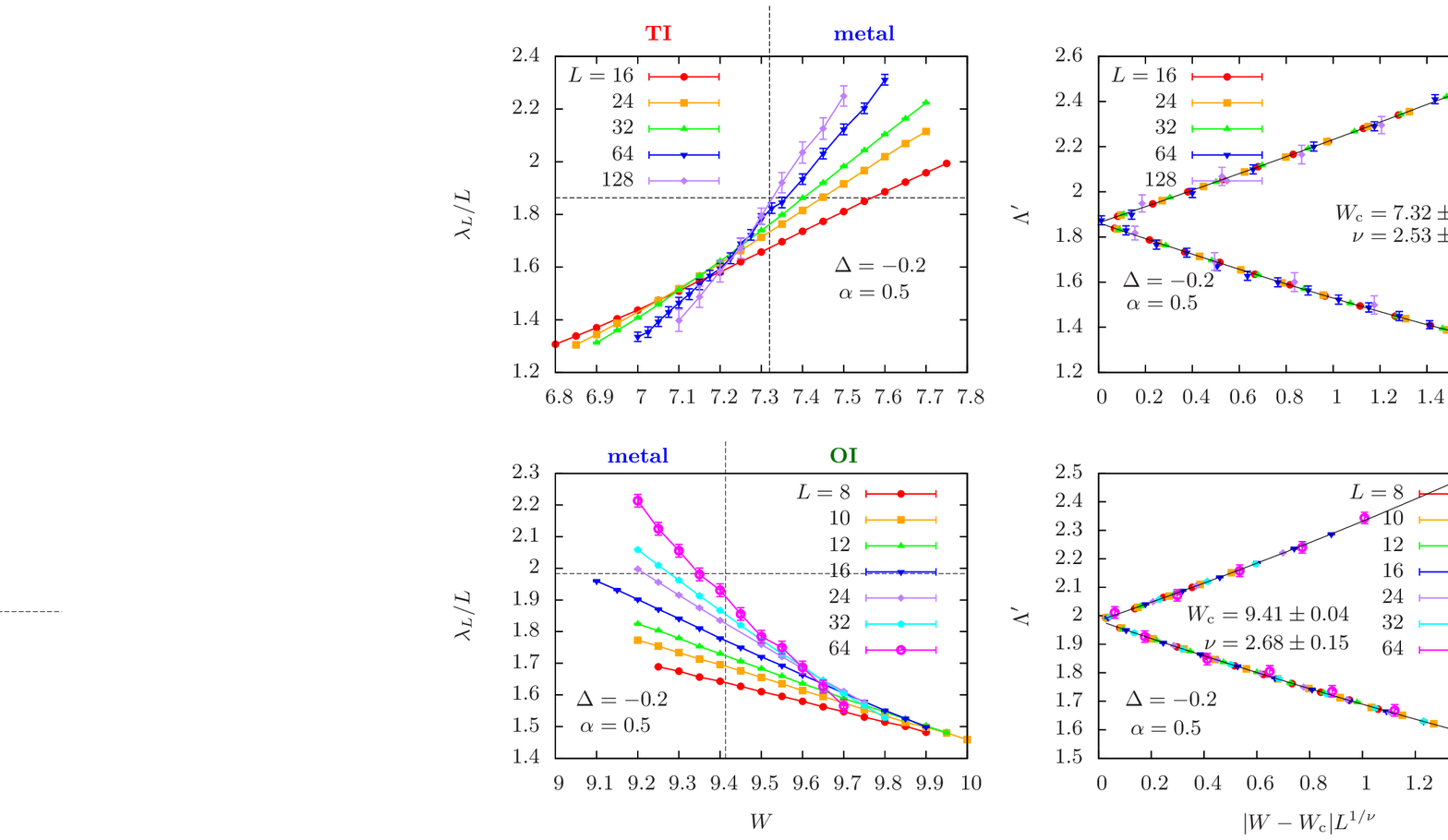}
\caption{The single-parameter scaling for $\Delta=-0.2$ and $\alpha=0.5$.}
\label{criticality_tai}
\end{figure*}

Now, we turn to discussion of the critical exponent of the metal-insulator transitions.
The largest localization length divided by the system width $\lambda_{L}/L$ and the single-parameter scaling part $\Lambda'$ in the vicinity of metal--topological (ordinary) insulator transition are shown in the upper (lower) panels of Fig \ref{critical_fig}.
In this calculation, the standard error of $\lambda_L/L$ less than 0.015 is adapted.
For $\Delta = 2$ and $\alpha=0.5$, the system is a TI in the clean limit.
As one increases disorder strength $W$, the system goes into metallic phase at $W \sim 5.93$ (the left-upper panel).
The corresponding  $\Lambda'$ is shown in the right-upper panel.
From the fitting, the critical exponent $\nu$ is determined as $\nu \sim 2.77 \pm 0.23$.
As shown in the lower panels of Fig. \ref{critical_fig},
for larger $W$, the system exhibits localization and turns into an ordinary insulator for $W > 9.78$.
Similarly to the previous case, from the fitting, $\nu$ is determined as $\nu \sim 2.60 \pm 0.06$.

Next, we discuss the criticality for $\Delta<0$. 
Here the system is a ordinary insulator for $W=0$ but shows the topological insulating state for a finite value of $W$.
The detailed phase structure and such a disorder--induced phase are discussed in the next section.
For $\Delta=-0.2$ (Fig. \ref{criticality_tai}),  from the single--parameter scaling analysis, it is confirmed that the system is in the disorder--induced TI phase for $W<7.32$, in metallic phase for $7.32<W<9.41$, and in ordinary insulator phase for $W>9.41$.
The critical exponent is evaluated for each transition as $\nu = 2.53 \pm 0.21$ and $\nu = 2.68 \pm 0.15$, respectively.
These values are consistent with those for $\Delta=2$.
Note that the finite--size effect for $\Delta = -0.2$ is much stronger than that for $\Delta=2$ since the energy gap $\sim |\Delta|$ is smaller.
Therefore, it is necessary to calculate the localization length in the large system up to $L=128$ ($L=64$) for the metal--topological(ordinary) insulator transition.

For both transitions from topological and ordinary insulators to metal, 
the critical value of localization length $\Lambda_0$ and the critical exponent $\nu$ are estimated to be $\Lambda_0 \sim 1.8$ and $\nu \sim 2.7$, which are consistent with those in symplectic systems, e.g., SU(2) model,\cite{asada02} the $\mathbb Z_2$ network models,\cite{obuse07,obuse08,obuse10,Koba} and the ${\mathbb Z_2}$ quantum kicked rotator.\cite{njeuwenburg12}
Namely, the topological non-triviality of the present system does not affect the quantum criticality of metal-insulator transition.
Our results are based on a tight--binding model of the actual systems, i.e., HgTe quantum well. 
Thus
it is proved that the critical phenomena of TI are described by the effective network models.
The obtained critical values and fitting parameters are summarized in Table \ref{critical_tab}.

\begin{table*}
\begin{tabular}{c||cc|cc}
\hline\hline
& \multicolumn{2}{c}{Metal--TI} & \multicolumn{2}{|c}{Metal--OI}
\\
$\Delta$ & 2.0 & $-0.2$ & $2.0$ & $-0.2$
\\
\hline
$\Lambda_0$ & $1.824 \pm 0.018$ & $1.86 \pm 0.08$ & $1.825 \pm 0.007$ & $1.98 \pm 0.1$ \hspace{0.5ex}
\\
$W_{\rm c}$ & $5.932 \pm 0.006$ & $7.32 \pm 0.03$ & $9.780 \pm 0.008$ & $9.41 \pm 0.04$
\\
$\nu$ & $2.77 \pm 0.23$ & $2.53 \pm 0.21$ & $2.60 \pm 0.06$ & $2.68 \pm 0.15$
\\
$y$ & $-1.57 \pm 0.37$ \hspace{1ex} & $-1.0 \pm 0.2$ \hspace{1ex} & $-3.21 \pm 0.74$ \hspace{1ex} & $-0.7 \pm 0.1$ \hspace{1ex}
\\
$N_{\rm d}$ & 69 & 70 & 59 & 86
\\
$\chi_{\rm r}^2$ & 0.88 & 1.0 & 1.02 & 1.0
\\
\hline\hline
\end{tabular}
\caption{Criticality of metal-topological insulator (Metal-TI) and metal-ordinary insulator (Metal-OI) transitions. 
Critical values of localization length divided by the system size $\Lambda_0$, disorder strength $W_{\rm c}$, the value of critical exponent $\nu$, irrelevant exponent $y$,
degrees of freedom $N_{\rm d}$, and reduced chi-square $\chi_{\rm r}^2$.
}
\label{critical_tab}
\end{table*}

\section{Phase diagram}
Together with the analyses on the critical exponent,
studying the structure of the phase diagram is an important aspect 
for highlighting the nature of the disordered TI and the disorder-induced TI (TAI).
A close comparison is made on
different situations 
(i) with or without carrier doping, and also 
(ii) with or without Rashba SOC.

\subsection{Phase diagram at $\Delta=-0.1$: TAI and symplectic metal phases}

\begin{figure}
\centering
\includegraphics{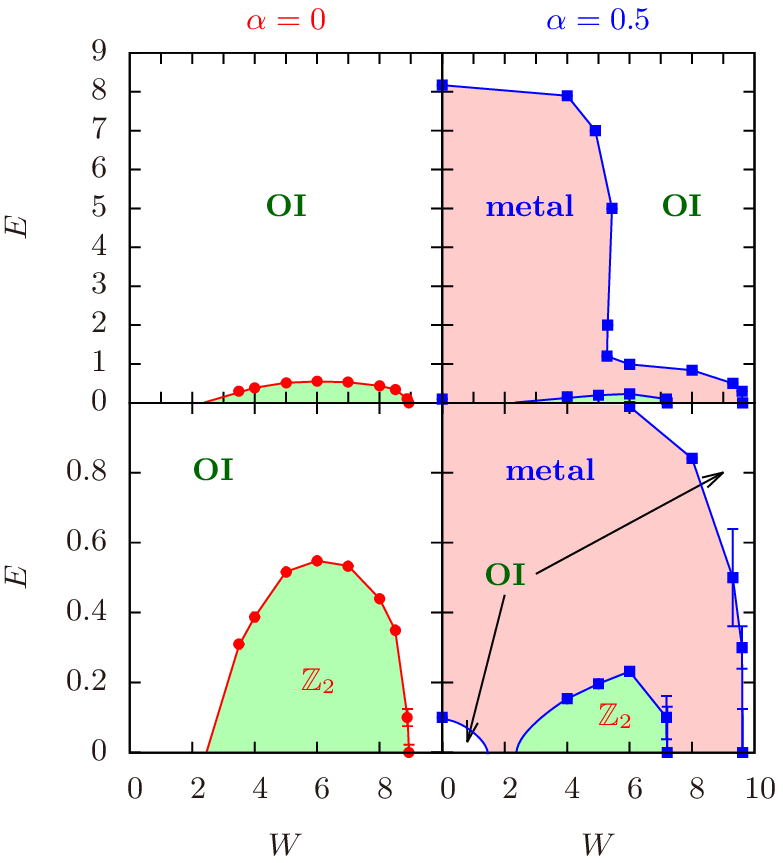}
\caption{
(Color online)
Phase diagram in $(W,E)$--plane at $\Delta = -0.1$.
The lower panel shows the enlargement of the upper panel.
The error bar denotes $\pm 2 \epsilon$, where $\epsilon$ is the standard error defined in Appendix \ref{method}.
}
\label{pd_Delta-01}
\end{figure}

Figure \ref{pd_Delta-01} shows the phase diagram for $\Delta=-0.1$, $\alpha=0$ (left) and $\alpha=0.5$ (right).
Transition lines between an ordinary and topological insulators for $\alpha=0$ and between a metal and insulator for $\alpha=0.5$ are denoted by circle (red line) and square (blue line) symbols, respectively.
In the clean limit ($W=0$), an ordinary insulator is realized for $\Delta=-0.1$.
The system for $\alpha=0$ belongs to the unitary class, where there is only the localized states except at the critical point.
For $\alpha=0$ (the left panels), 
a TI is realized 
in the isolated region ($2 < W < 9$) in $(W,E)$--plane, i.e., disorder induces a TAI, which is consistent with Refs. \onlinecite{li09,groth09}.
Even if one introduce Rashba SOC $\alpha$, 
the TAI can survive as shown in the right-lower panel of Fig. \ref{pd_Delta-01}, where there exists the TAI region in $2 < W < 7$, although it is reduced as compared with that for $\alpha=0$ (the left-lower panel).

In addition to this,
in the presence of Rashba SOC,
a metallic phase can appear due to anti-localization.\cite{hikami80,ando89}
Actually, metallic phase 
 spreads over a smaller $W$ ($<10$) and $E$ ($<9$) area, as shown in the right-upper panel of Fig. \ref{pd_Delta-01}.
As one decreases $\alpha$, metallic region shrinks. 
And then it vanishes in the case of $\alpha=0$, except for the transition line between topological and ordinary insulators.
The transition lines between metal and ordinary insulator for $\alpha=0.5$ reach to $E \sim 0.1$ and $E \sim 8$ at $W=0$.
These two points are the band edge for $W=0$ (See Fig. \ref{dos} in Appendix \ref{escleanlimit}), i.e., metal-insulator transition point in the clean limit.

\subsection{Phase diagram at $E=0$: nature of the TAI phase}

\begin{figure}
\centering
\includegraphics{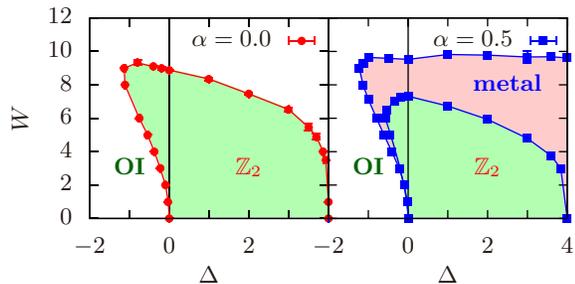}
\caption{
(Color online)
Phase diagram in $(\Delta, W)$--plane at $E=0$.
}
\label{pd_e00}
\end{figure}

Phase diagram at $E=0$ more precise than that in Ref. \onlinecite{yamakage11} is shown in Fig. \ref{pd_e00}.
It is shown only for $\Delta < 4$ since it is symmetric with respect to $\Delta=4$ (Appendix \ref{phs}).
A finite metallic region is
found for $\alpha=0.5$, which partitions the two topologically distinct insulating phases.
The inner (week disorder) region is a topological insulating phase, and the outer (strong disorder) region is an ordinary insulating phase.
In the region of $-1 < \Delta < 0$, $W<10$ for $\alpha=0$, and of $-0.5<\Delta<0$, $W<7$ for $\alpha=0.5$, the TAI is realized.
The TAI is continuously connected to the clean TI ($0<\Delta<4$ and $W=0$),
i.e., both TI and TAI are essentially the same.\cite{yamakage11,prodan11_2,prodan11}

The transition line between the metal and ordinary insulator is located roughly at $W=10$, which is nearly equal to the band width and independent of $\Delta$. 
Since Anderson localization occurs when the energy scale of disorder is larger than that of band width,
metallic phase cannot exist for $W>10$ even if the Rashba SOC becomes stronger.
It is expected that
as one increases $\alpha$ from $\alpha=0$,
the metallic region gradually spreads around the transition line for $\alpha=0$ and it converges roughly to $W<10$.

The inner transition line located between metallic  and topological insulating phases, on the other hand, depend on not only band width but also detailed band structure, i.e., $\alpha$ and $\Delta$.
As one approaches $\Delta=4$, where the band gap closes at $W=0$, the critical disorder strength $W_{\rm c}$ decreases and the transition line connects to $(\Delta, W) = (4,0)$. 
TI phases in $0<\Delta<4$ and $4<\Delta<8$ are separated from each other, even in the presence of disorder.
The TIs for $0<\Delta<4$ and $4<\Delta<8$ have the same strong topological number $\nu=1$.
However, for $4 < \Delta < 8$,
bound states appear at a dislocation,\cite{zaanen12, matsumoto} while it does not appear for $0<\Delta<4$.
In this sense, these two topological phases are distinguished from each other.

\subsection{Phase diagram at $E=0.5$: Carrier-doping effects}

\begin{figure}
\centering
\includegraphics{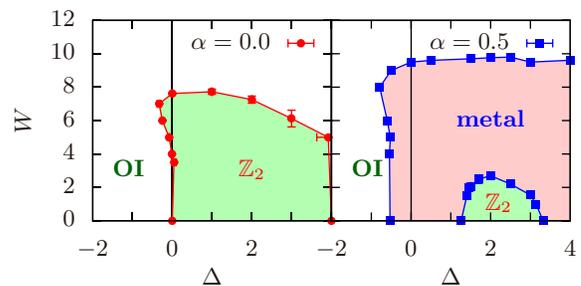}
\caption{
(Color online)
Phase diagram in $(\Delta, W)$--plane at $E=0.5$.
}
\label{pd_e05}
\end{figure}

Next, we discuss carrier-doping effects in the disordered system.
Figure \ref{pd_e05} shows the phase diagram for $E=0.5$, which corresponds to the case of Fig. \ref{pd_e00} with carrier doping.
In the absence of Rashba SOC ($\alpha=0$), Anderson localization always occurs except on the critical point, therefore carrier-doping does not qualitatively change the phase structure, as shown in the left panel of Fig. \ref{pd_e05}.
The TI phase region slightly shrinks by carrier--doping.
In the clean limit, only the sign of $\Delta$ is relevant to the topological phase transition. 
Thus the critical point is given by $\Delta = 0$ for $W \sim 0$.
In contrast,
the phase structure for $\alpha=0.5$ (the right panel of Fig. \ref{pd_e05}) in the weakly disordered region drastically changes from that for $E=0$ (Fig. \ref{pd_e00}).
Since a metallic phase can appear due to anti-localization, the critical lines of the metal-insulator transitions connect to those for $W=0$ ($\Delta=-0.5$ and $\Delta = 1.2$, see Appendix \ref{pdcl}).
As a result, a wider metallic region is realized in the weakly disordered regime ($W<6$) in the presence of Rashba SOC.
The TAI region is dominated by the metallic phase.

\subsection{Limitations of the SCBA picture}

\begin{figure*}
\centering
\includegraphics{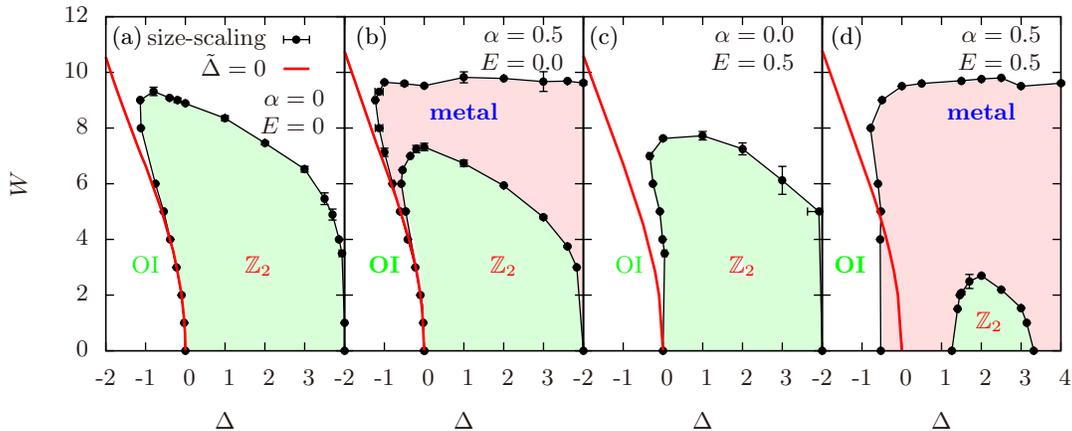}
\caption{
Comparison between the critical points calculated by the size-scaling and by the SCBA for $\alpha=0$, $E=0$ (a), $\alpha=0.5$, $E=0$ (b), $\alpha=0$, $E=0.5$ (c), and $\alpha=0.5$, $E=0.5$ (d).
The band inversion points ($\tilde \Delta = 0$) calculated by the SCBA are denoted by the (red) solid line.
The critical points determined by the size-scaling are denoted by the circle symbols, which are the same as in Fig. \ref{pd_e00} and \ref{pd_e05}.
}
\label{scba}
\end{figure*}

A possible interpretation of the TAI behavior is based on
the observation that in Dirac systems 
disorder not only renormalizes the magnitude of the mass (i.e., the band gap) 
but sometimes even changes its sign.\cite{groth09}
Thus, by reversing the sign of $\Delta$ from
a negative to a positive value,
disorder 
by itself can
convert an ordinary insulator to a TI.
The self-consistent Born approximation (SCBA)
is a simple and useful method for quantifying the above mass renormalization.
\cite{shindou09,groth09}
Here, we remark,
on contrary to the previous studies, that 
upon carrier doping
SCBA is no longer a quantitatively good approximation.

The self-energy in the disordered system is given by
\begin{align}
 \Sigma(E) = \frac{W^2}{12}\int \frac{d^2k}{(2\pi)^2} G_{\bm k}(E),
\end{align}
within SCBA, where $G_{\bm k}(E) = (E-H_0(\bm k)-\Sigma(E)  + i0)^{-1}$ is the Green's function, and $H_0(\bm k)$ is the Hamiltonian in the clean system.
We iteratively solve the above self-consistent equation for $\Sigma(E)$.
The obtained self-energy $\Sigma(E)$ is decomposed into $\Sigma(E) = \Sigma_0(E) \sigma_0 s_0 + \Sigma_z(E) \sigma_z s_0$ due to the symmetry of system.
Note that terms proportional to $\sigma_0 s_0$ and $\sigma_z s_0$ are independent of $\bm k$.
The Renormalized mass $\tilde \Delta$ and Fermi level $\tilde E$ are given by
\begin{align}
 \tilde \Delta(E) &= \Delta + \mathrm{Re}\,\Sigma_z(E),
 \\
 \tilde E(E) &= E - \mathrm{Re}\,\Sigma_0(E).
\end{align}
In the renormalization picture, a TI is realized for $\tilde \Delta(E) > 0$.

In the following, we discuss the phase boundary in the carrier-doped ($E=0.5$) and undoped ($E=0$) cases.
For $\Delta < 0$, as one increases $W$, $\Sigma_z(E)$ increases and $\tilde \Delta(E)$  becomes zero for a certain value of $W$ ($W=W_{\rm c}$) and $\tilde \Delta(E)$ becomes positive for $W>W_{\rm c}$, where the system is a TI.
The band inversion point $\tilde \Delta(E)=0$, i.e., $W=W_{\rm c}$,
is denoted by the (red) solid line in Fig. \ref{scba}. 
In the lightly doped cases [$E=0$, Figs. \ref{scba}(a) and \ref{scba}(b)], the SCBA gives quantitatively good result in weakly disordered region ($W<5$).
For $\alpha=0$ [Fig. \ref{scba}(a)], the band inversion points obtained by the SCBA coincides with the critical points of the topological phase transition.
And also, for $\alpha=0.5$ [Fig. \ref{scba}(b)], the band inversion points coincide with the critical points of the TAI about for $W<4$.
In the weakly disordered regime, metallic phase is realized only in a narrow region since the system has no carrier in the clean limit. 
Here anti-localization effect is not important thus the SCBA and  renormalization picture still work well.

In contrast, in the carrier-doped cases of $E=0.5$ [Figs. \ref{scba}(c) and \ref{scba}(d)], the SCBA cannot quantitatively give the critical points of the topological phase transition.
The band inversion point of $\tilde \Delta=0$ does not coincide with the actual critical points determined by the size-scaling.
This mismatch suggests that Anderson localization, which is beyond the range of SCBA, plays an important role for the topological quantum phase transition for carrier-doped cases.
For the carrier-undoped case, the system is already insulating in the clean limit and does not directly related to the localization.
In such cases, SCBA gives the correct critical points.
For $\alpha = 0.5$ [Fig. \ref{scba}(d)], metallic phase widely appears,  
dominating
the TAI region. 
Thus the mass renormalization picture becomes invalid.

\section{Summary}
In this paper, we have clarified the criticality of the metal--insulator transitions in a disordered two-dimensional TI.
The obtained critical exponent is $\nu \sim 2.7$ for both the critical points of metal--ordinary and metal--topological insulators. 
Moreover, the critical exponent of metal--TAI transition is also estimated to be the same value.
This criticality is consistent with that in the symplectic class.
Namely, topological non-triviality does not affect criticality of the phase transition.
And also, 
Hamiltonian based on the actual system is shown to be safely mapped to the effective network model.

The phase diagram of the disordered system has also been studied.
We have confirmed that the ``TAI" is not a distinct phase,  employing the single-parameter scaling.
The TAI is continuously connected to the TI in the clean limit, without any phase transition.
The critical exponent of metal--TAI transition is also not a distinct one.

In addition,
carrier-doping and Rashba SOC yield a wide metallic region in the phase diagram, due to the anti-localization.
The TAI is affected by carrier-doping:  
the renormalized mass $\tilde \Delta(E)$ does not quantitatively give the critical points between the ordinary insulator and TAI.

A two-dimensional TI implemented in a quantum well has advantages for experiments. 
The Fermi level and strength of Rashba SOC can be tuned by the gating.
Moreover, it is possible to tune Rashba SOC with tuning  structural-inversion-asymmetry of the quantum well.
We believe that our results obtained in the present paper will be verified in such quantum wells with HgTe\cite{konig07} and InAs.\cite{liu08,knez10,knez11,knez12}

\begin{acknowledgments}
AY acknowledges Grant-in-Aid for JSPS Fellows under
Grants No. 08J56061. 
AY and KI are supported by
the ``Topological Quantum Phenomena" [Nos. 22103005 (AY) and 23103511 (KI)] Grant-in Aid for Scientific Research on Innovative Areas from the Ministry of Education, Culture, Sports, Science and Technology (MEXT) of Japan.
KN and KI are supported by
Grant-in-Aid for Young Scientists (B) under Grants Nos.
24740211 (KN) and 19740189 (KI). 
KN is also supported by FIRST program (JSPS).
AY is grateful to  W. Izumida for discussion about the parallel computation.
\end{acknowledgments}


\appendix

\section{Determination of the matrix elements 
of
Rashba SOC: symmetry considerations}

\label{rashba}


The model
Hamiltonian for the HgTe quantum well with inversion symmetry has been derived
in Ref. \cite{bernevig06}
Here, we extend this effective
model by taking into account the Rashba SOC.
The explicit form of the Rashba SOC is determined by symmetry considerations.
The system we consider has a $C_{4v}$ symmetry,
and is modeled as a tight-binding model defined on a square lattice.
In the following, we identify non-vanishing matrix elements of the SOC
allowed by this symmetry
that are also associated with an electric field perpendicular to the $xy$-plane.

In our tight-binding description
the internal degrees of freedom of the system is labeled by $z$-component 
of total angular momentum $j_z=\pm 1/2$ for $s$-orbital and $j_z = \pm3/2$ for $X\pm iY$-orbital.
The basis is taken as $(|1/2 \rangle, | 3/2 \rangle, \left|-1/2 \right\rangle, \left| -3/2 \right\rangle)$.
The system has time-reversal $\Theta$, four-fold rotational $R_4$, and mirror $M_y$ symmetries.
The corresponding (anti-)unitary matrices acting on the internal degrees of freedom are given by
\begin{align}
 \Theta &= -i s_y \mathcal K,
 \\
 R_4 &= e^{-i j_z \pi/2} = \frac{1}{\sqrt 2} \sigma_z - \frac{i}{\sqrt 2} s_z,
 \\
 M_y &=  s_y,
\end{align}
where $\sigma_i$ and $s_i$ are Pauli matrices in the orbital and spin spaces, respectively.
The on-site potential $\epsilon_{\bm r}$ should satisfy the following relations.
\begin{align}
 \Theta \epsilon_{\bm r} \Theta^{-1} 
&= \epsilon_{\bm r},
 \\
 R_4 \epsilon_{\bm r} R_4^\dag 
&= \epsilon_{\bm r},
\\
M_y \epsilon_{\bm r} M_y^\dag 
&= \epsilon_{\bm r}.
\end{align}
This restricts the form of $\epsilon_{\bm r}$ to 
\begin{align}
 \epsilon_{\bm r} = (\Delta-4B) \sigma_z s_0,
\label{epsilon_r}
\end{align}
where the parameters $\Delta$ and $B$ are real numbers, 
Similarly, $t_x$ has the following relations,
\begin{align}
 \Theta t_x \Theta^{-1} 
&= t_x,
\\
R_4^2 t_x R_4^{2\dag} 
&= t_x^\dag,
\label{r42tx}
\\
M_y t_x M_y^\dag 
&= t_x,
\end{align}
with $R_4^2 = \sigma_z s_z$.
Thus,
the form of $t_x$ is determined as
\begin{align}
 t_x &= D \sigma_0 s_0 + B \sigma_z s_0 
-i \frac{A}{2} \sigma_y s_0 + i \frac{\alpha}{2} \sigma_z s_y 
\nonumber \\ & \qquad
+ i\frac{\alpha'}{2} \sigma_0 s_y + \frac{\alpha_{\rm O}}{2} \sigma_y s_y,
\label{tx}
\end{align}
where all the parameters $A$, $D$, and $\alpha$ are real numbers. 
$\alpha$, $\alpha'$, and $\alpha_{\rm O}$  denote strengths of Rashba SOCs.
$t_y$ is obtained by the rotation of $t_x$ as
\begin{align}
 t_y &= R_4 t_x R_4^\dag
 \nonumber \\
 &= D \sigma_0 s_0 + B \sigma_z s_0 
- i \frac{A}{2} \sigma_x s_z 
- i \frac{\alpha}{2} \sigma_0 s_x
 \nonumber\\ & \qquad
 - i\frac{\alpha'}{2} \sigma_z s_x - \frac{\alpha_{\rm O}}{2} \sigma_y s_y.
 \label{ty}
\end{align}

Here, we confirm that Rashba SOC does not appear in inversion-symmetric systems.
If we consider $D_{4h}$ symmetry, two-fold rotation $R_2=e^{-i j_x \pi}=s_x$ along the $x$-axis also becomes a symmetric operation. Consequently, the following relation holds.
\begin{align}
 R_2 t_x R_2^\dag = t_x.
\end{align}
This leads to $\alpha=\alpha'=\alpha_{\rm O}=0$.
To be sure, Rashba SOC is not allowed in inversion-symmetric systems.

In the momentum space, the Hamiltonian with Rashba SOC reads
\begin{align}
 H(\bm k) 
 &= 
 (\Delta - 4 B) \sigma_z s_0 
\nonumber\\ & \qquad
+
 2(D \sigma_0 + B \sigma_z) s_0 (\cos k_x + \cos k_y)
 \nonumber\\ & \qquad
 + [A \sigma_y s_0 - (\alpha \sigma_z + \alpha' \sigma_0) s_y] \sin k_x
 \nonumber\\ & \qquad
 + [A \sigma_x s_z + (\alpha \sigma_0 + \alpha' \sigma_z) s_x] \sin k_y
 \nonumber\\ & \qquad
 + \alpha_{\rm O} \sigma_y s_y (\cos k_x - \cos k_y).
 \label{Hk}
\end{align}
In the above expression, the linear terms of $k$ near $\Gamma$-point with spin flip 
[$-(\alpha \sigma_z + \alpha' \sigma_0) s_y \sin k_x + (\alpha \sigma_0 + \alpha' \sigma_z) s_x \sin k_y$] 
are present, differently from the case in Ref. \onlinecite{rothe10}.
This is because the symmetry in a square lattice ($C_{4v}$) is lower than  that in $\bm k \cdot \bm p$ theory ($C_{\infty v}$).
Actually,  in the axial symmetric case ($C_{\infty v}$), 
the spin-flip term for $j_z=\pm 3/2$ states satisfies
\begin{align}
\langle \bm k, 3/2 | H | \bm k, -3/2 \rangle 
 = e^{i 3 \theta} \langle \bm k', 3/2 | H | \bm k', -3/2 \rangle,
\end{align}
where $\bm k' = e^{-i j_z \theta} \bm k$.
As a result, the leading term is
\begin{align}
\langle \bm k, 3/2 | H | \bm k, -3/2 \rangle 
\propto k_{-}^3,
\end{align}
namely, the linear term is prohibited.
The linear term is
possible in the discrete symmetric case $C_{4v}$.
In the actual calculation, we set $D=\alpha' = \alpha_{\rm O} = 0$.

\section{Remarks on the clean limit}
\label{cleanlimit}


\subsection{
Electronic states in the clean limit
}
\label{escleanlimit}

\begin{figure}
\centering
\includegraphics[scale=0.66]{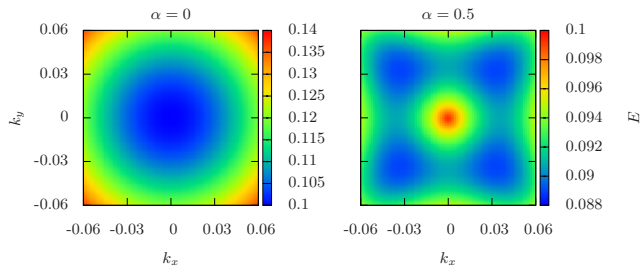}
\caption{(Color online) Energy spectra of the lower conduction band for $\alpha=0$ (left) and $\alpha=0.5$ (right).
The parameters are given as $A=B=1$.}
\label{energyband}
\end{figure}

FIG. \ref{energyband} shows energy spectra of the conduction band for $\alpha=0$ and $\alpha=0.5$ in the clean limit $W=0$.
For $\alpha=0$, the bottom of the conduction band
is located at $\Gamma$ point.
When $\alpha$ is switched on, spin-degeneracy is lifted and the minima of the energy band become located at finite $\bm k$-points as shown in the right panel of Fig. \ref{energyband}.

\begin{figure}
\centering
\includegraphics[scale=1]{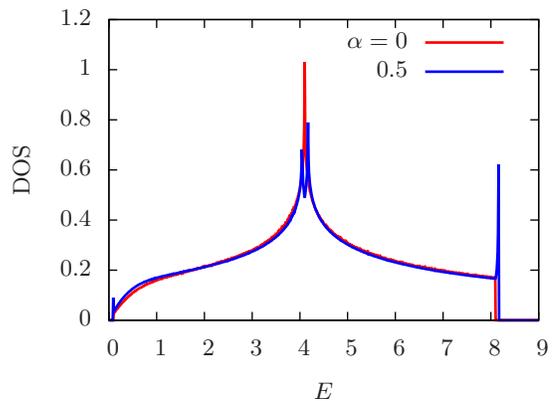}
\caption{(Color online) Density of states in the presence/absence of Rashba SOC $\alpha$.}
\label{dos}
\end{figure}

The corresponding density of states is shown in Fig. \ref{dos}, which is shown only for $E>0$ since it is symmetric with respect to $E=0$, due to the chiral symmetry.
The behavior is consistent with that in Ref \onlinecite{ando89}, where the $s$-orbital is focused on and the $p$-orbitals are integrated out.
The conduction band exists about in $0.1 < E < 8$.
In the presence of Rashba SOC,
the density of states takes a larger value at the band edge ($E \sim 0.1$ and $E \sim 8$), due to the multiple-minima of the conduction band.
In the continuum limit, the density of states at the band edge diverges since the shape of energy band becomes  wine--bottle.

\subsection{
Phase diagram in the clean limit
} 

\label{pdcl}

When $s_z$ is conserved ($\alpha=0$), TI phase with a quantized spin Hall conductance is realized.
It is known that such a topologically non-trivial phase is robust against weak Rashba SOC, which breaks spin rotational symmetry, i.e., $s_z$ is not conserved.
In this case, although the spin Hall conductance is not quantized, 
we can define the $\mathbb Z_2$ topological invariant
as discussed by Kane, Mele and Fu.\cite{kane05_2,fu06} 
However, as one increases Rashba SOC, the magnitude of the band gap decreases and vanishes at the critical value $\alpha = \alpha_{\rm c}$. 
For  $\alpha \geq \alpha_{\rm c}$, the system becomes metallic and irrelevant for topological quantum phenomena.

\begin{figure}
\centering
\includegraphics{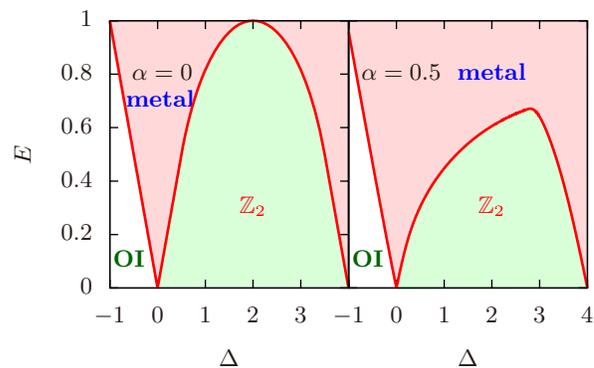}
\caption{(Color online) Energy ($E$) -- mass ($\Delta$) phase diagrams for $\alpha=0$ and $\alpha=0.5$ in the clean system. 
$\mathbb Z_2$ and OI denote $\mathbb Z_2$ TI and ordinary insulator respectively. 
The parameters are set to $A=B=1$.}
\label{clean}
\end{figure}

Figure \ref{clean} shows the phase diagram in the $(\Delta, E)$--plane determined by evaluating the magnitude of the band gap.
In the absence of $\alpha$,\cite{bernevig06,imura10} a TI is realized in $0<\Delta<4B$ with $\sigma_{xy}=e/2\pi$ and $4B<\Delta<8B$ with $\sigma_{xy}=-e/2\pi$, where $\sigma_{xy}$ is a quantized spin Hall conductance with respect to $z$-component of spin.
At $\Delta=0$, $4B$, and $8B$, the band gap vanishes (zero-gap semiconductor) and the phase transition occurs. 
An ordinary insulator is realized in $\Delta<0$ and $\Delta>8B$.
In the presence of $\alpha$,
the gapless points $\Delta=0, 4B$, and $8B$ still remain, and
a TI phase can exist about in $E < 0.5$ for $A=B=1$ and $\alpha=0.5$, as shown in Fig. \ref{clean}.
If one further increases $\alpha$, the region of TI phase shrinks, and vanishes at $\alpha \geq \alpha_{\rm c} = A$.
Moreover,
the phase diagram is symmetric with respect to $\Delta=4$ due to the symmetry discussed in  Appendix \ref{phs}.

\section{The transfer matrix and related numerical methods}
\label{method}

The transfer matrix method is used
for estimating numerically the localization length of the system.
Here, we describe some details of the method,
and the protocol of our numerical simulation.
In addition, we show the definition of the standard error.

Let us consider the following Schr\"odinger equation
discretized on a {two-dimensional}
lattice:
\begin{align}
 E \psi_{I,J} &= \epsilon_{I,J} \psi_{I,J} + t_x \psi_{I+1,J} + t_x^\dag \psi_{I-1,J} 
\nonumber\\ & \quad
+ t_y \psi_{I,J+1} + t^\dag_y \psi_{I,J-1}.
\end{align}
This can be rearranged into a {one-dimensional} form in terms of
the transfer matrix $M_{I}$ as
\begin{align}
\begin{pmatrix}
  \psi_{I+1}
  \\
  \psi_I
\end{pmatrix}
= 
M_I
\begin{pmatrix}
  \psi_{I}
  \\
  \psi_{I-1}
\end{pmatrix},
\end{align}
where
$\psi_I = (\psi_{I,1}, \cdots \psi_{I,L})^{\rm T}$.
$M_{I}$ is generally a $2gL$-dimensional matrix
with $g$ being the number of internal degrees of freedom
($g=4$ in the present model),
and is given by
\begin{align}
 M_I = 
\begin{pmatrix}
t_x^{-1} (E- \epsilon_I) - t_x^{-1} t_y l_+ - t_x^{-1} t_y^\dag l_- 
&
- t_x^{-1} t_x^\dag
\\
1 & 0
\end{pmatrix},
\end{align}
where $\epsilon_I = \mathrm{diag} (\epsilon_{I,1}, \cdots, \epsilon_{I,L})$, $(l_+)_{I,J} = \delta_{I+1,J}$, $l_- = l_+^\dag$.
Due to the disorder, wave function decays exponentially
as $\psi_i \sim e^{\pm I/\lambda_i}$, where $\lambda_i$ is a localization length.
$\lambda_i$ is deduced from the transfer matrix as
\begin{align}
 \frac{\lambda_i}{L} = \left( \frac{L}{L_x} \ln |m_i| \right)^{-1}, (i=1,\cdots, 2gL)
\end{align}
where $m_i$ is the eigenvalue of $M^{(L_x)} = M_{L_x-1} \cdots M_1$.

Thus, to estimate $\lambda_i$
we have only to calculate the eigenvalues of 
the product of transfer matrices, $M^{(L_x)} = \prod_{I=1}^{L_x} M_I$.
A direct calculation, however, usually fails,
since numerical error tends to accumulate in the product of matrices.
A common resort to this failure is the use of 
$QR$-decomposition,
{\it i.e.},
after an iterative use of the decomposition as
\begin{align}
 M_1 &= Q_1 R_1,
 \\
 M_{I+1} Q_I &= Q_{I+1} R_{I+1},
\end{align}
one finds
\begin{align}
\label{mlx}
 M^{(L_x)} = M_{L_x} \cdots M_{1} = Q_{L_x} R_{L_x} \cdots R_1,
\end{align}
where
$Q_I$'s and $R_I$'s
are respectively unitary and trigonal matrices.
Neglecting $Q_{L_x}$ in the above equation
turns out to be a fairly good approximation 
for large enough $L_x$.
Then, $M^{(L_x)}$ 
can be regarded as a product of trigonal matrices.
This means that one can safely estimate the eigenvalues of this matrix 
simply by multiplying the corresponding diagonal element of 
each trigonal matrix $R_I$.
Recall that the eigenvalues of a trigonal matrix are its diagonal elements,
and a product of such trigonal matrices are also trigonal.

The actual calculation has been done by following the steps as listed below:
\cite{markos06}

\begin{itemize}
\item[i.]{
Set the initial value of the unitary matrix $Q$,
{\it e.g.},
to be $\bm 1$, for simplicity.
Set also as $\bm d = (0, \cdots, 0), \bm e = (0, \cdots, 0), L_x=0$.
}
\item[ii.]{
Generate the $n_i$ transfer matrices and calculate the product $\tilde M=M_{n_i} \cdots M_{1} Q$.
}
\item[iii.]{
QR-decomposition. $\tilde M = Q R$.
}
\item[iv.]{
Store the diagonal components of $R$ as $d_a = d_a + \ln |r_a|, e_a = e_a + (\ln |r_a|)^2$, where $r_a$ is the $a$-th diagonal component.
Calculate the system length $L_x$: $L_x = L_x + n_i$.
}
\item[v.]{
The localization length and the approximated standard error $\epsilon_a$ are given by $\lambda_a/L = L_x/(d_a L)$ and $\epsilon_a = L(\lambda_a/L)^2 \sqrt{\epsilon_a/L_x^2 - (d_a/L_x)^2 n_i/L_x}$ respectively. 
}
\item[vi.]{
If the error $\epsilon_a$ is larger than the accuracy goal $\epsilon_0$, go to step ii.
}
\end{itemize}
In step iv. we have introduced a parameter $n_i$.
This must be chosen appropriately, in accordance with the nature of the problem.
If $n_i$ is too large, 
the resulting numerical error would become intolerable.
An appropriate choice of $n_i$ can be made by tuning
$d_a$ to be symmetric, 
{\it i.e.},
$d_a$ should appear as a pair such that $d_a = -d_b$.
This comes from the fact that
the eigenvalues also appear as pairs $e^{\pm L_x/\lambda_1}, \cdots, e^{\pm L_x/\lambda_L}$.
Here, we have chosen as $n_i=5$.
{
and have checked 
that 
the symmetry of $d_a = -d_b$ is satisfied as $(d_a+d_{b})/(d_a-d_b) < 0.02$.}

Last but not the least,
we have also employed a parallel computation.
Using $N_{\rm CPU}=64$ CPUs, 
we calculate the localization length
of $N_{\rm CPU}$ independent systems
with the accuracy of $\epsilon_L^{(i)} \sim \epsilon_0 \times (N_{\rm CPU})^{1/2}$ 
in each calculating node ($i=1, \cdots, N_{\rm CPU}$). 
Collecting the results as 
$L_x = \sum_{i=1}^{N_{\rm CPU}} L_x^{(i)}, \ 
d_a = \sum_{i=1}^{N_{\rm CPU}} d_a^{(i)}, \ 
e_a = \sum_{i=1}^{N_{\rm CPU}} e_a^{(i)}$, 
we obtain the average $\lambda_L/L$ and the error $\epsilon_L$.
This parallelization is actually not that trivial,
since the $j$-th ``sampling'' 
does depend on $(j-1)$-th sampling by the unitary matrix $Q$.
Yet, we believe that
the parallelization is fully justified
for sufficiently large $L_x^{i}/n_i$.

\begin{figure*}
\begin{minipage}{0.49\hsize}
\centering
\includegraphics{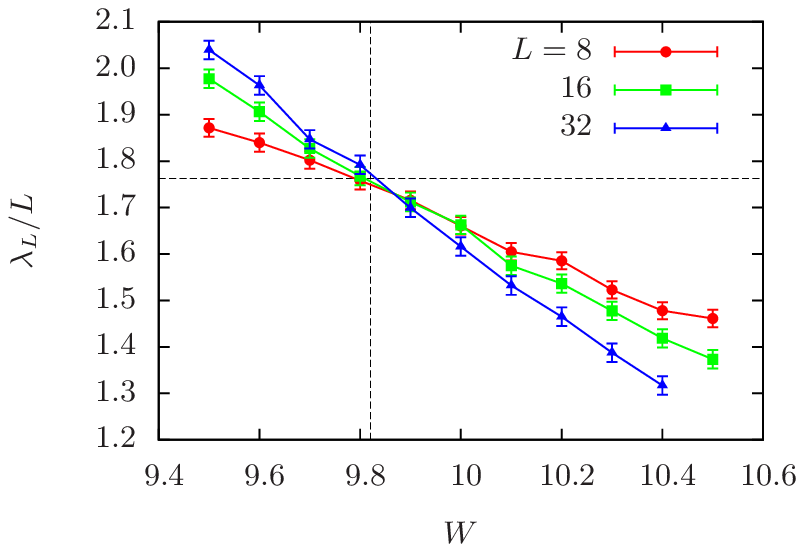}
\end{minipage}
\begin{minipage}{0.49\hsize}
\centering
\includegraphics{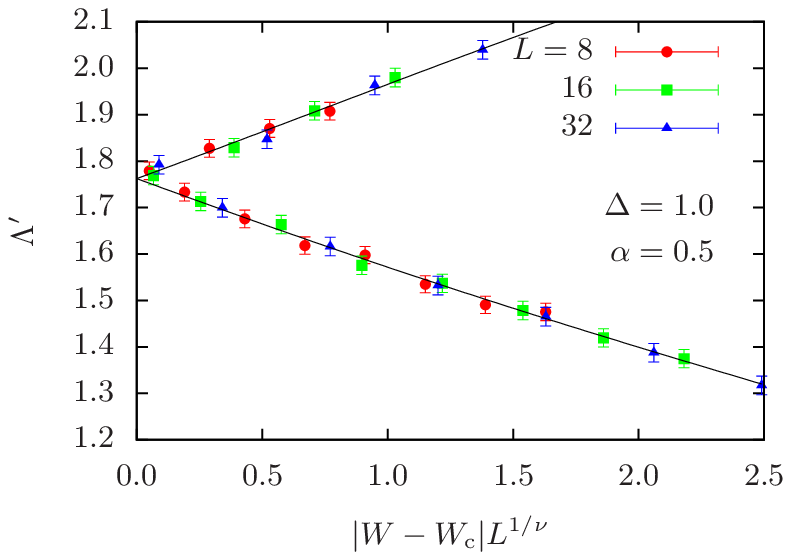}
\end{minipage}
\caption{The localization length and its single--parameter scaling in the presence of Rashba spin--orbit interaction ($\alpha = 0.5$) are shown at $\Delta=1$ and $E=0$.
The other parameters are taken as $A=B=1$.
Periodic boundary condition is applied.
The critical point  $(W_{\rm c}, \Lambda_0) = (9.821, 1.762)$ is denoted by the dashed line.}
\label{ll2}
\end{figure*} 
\begin{figure*}
\centering
\begin{minipage}{0.49\hsize}
\includegraphics{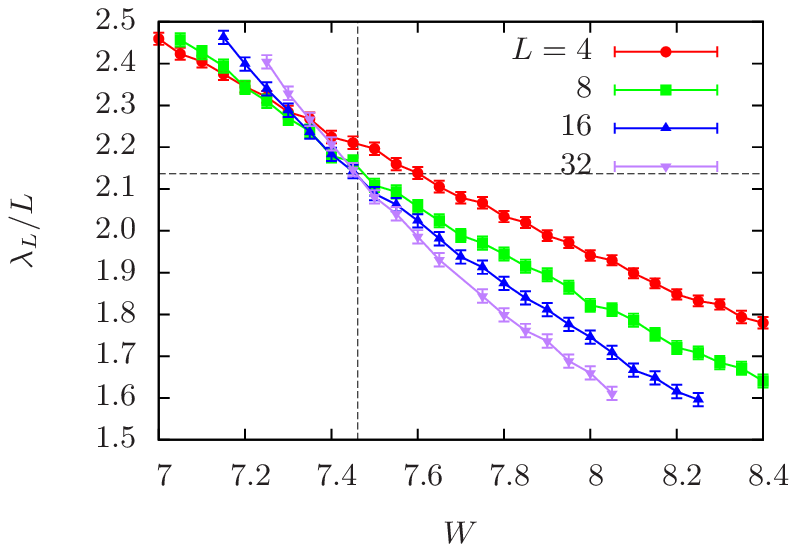}
\end{minipage}
\begin{minipage}{0.49\hsize}
\centering
\includegraphics{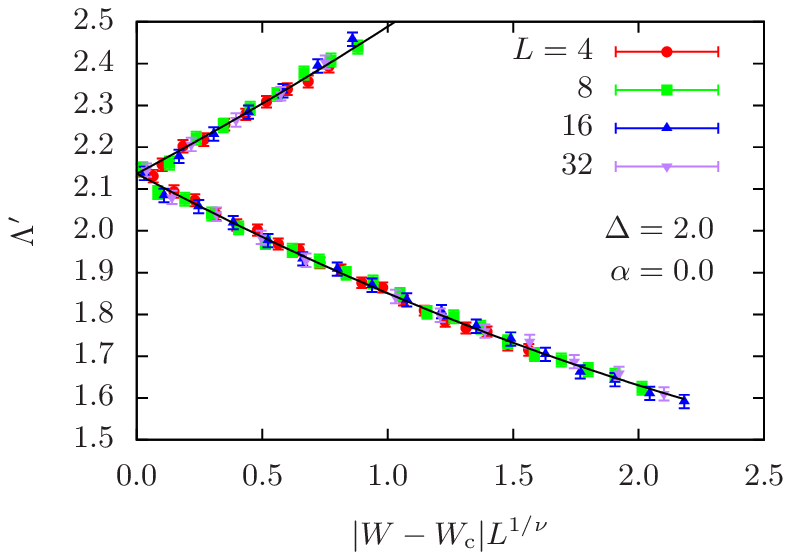}
\end{minipage}
\caption{The localization length and its single--parameter scaling in the absence of Rashba spin--orbit interaction ($\alpha=0$).
Open boundary condition is applied.
The critical point $(W_{\rm c}, \Lambda_0) = (7.461, 2.137)$ is denoted by the dashed line. The other parameters are taken as in Fig. \ref{ll2}.}
\label{ll1}
\end{figure*}

\section{Remarks on the choice of boundary conditions: determining the critical point}
\label{remarks}

In the actual calculation finite-size effects
can be a serious problem for determining the critical point.
To reduce the size effects
it is sometimes useful to apply an open boundary condition,
{\it e.g.},
in the case of insulator-insulator transitions,
and in the regime of weak disorder.

\subsection{Periodic boundary condition}

Two panels of Fig. \ref{ll2} represent a typical example of our size-scaling analysis.
The left panel shows the localization length (divided by $L$) as a function of
disorder strength as varying the size $L$ of the system.
To perform numerical estimate of the localization length
we make the system in the form of a long, quasi-one dimensional tube 
of length $L_x \simeq \infty$ and circumference $L$
({\it periodic boundary condition} is applied in the circumferential direction).
Here, the Rashba SOC is finite ($\alpha=0.5$), 
and the phase diagram shows a metallic region 
between the two -- one topological, the other ordinary -- insulating phases
(the system belongs to the symplectic symmetry class).
Furthermore, in the range of $W$ relevant in these plots,
the system undergoes a quantum phase transition 
from a metal to an ordinary insulator.
This appears, in the left panel of Fig. \ref{ll2},
as the change of the size-dependence of $\lambda_L /L$
from an increasing to a decreasing behavior,
with a size-independent critical point in the middle,
located roughly around $W\sim 9.8$.
More precise determination of the critical point needs,
however, single-parameter scaling analysis
as shown in the right panel of Fig. \ref{ll2},
due to a non-negligible finite-size effect.
Dashed lines in the left panel indicate the position
of such a critical point:
$(W_{\rm c}, \Lambda_0) = (9.82 \pm 0.20, 1.76 \pm 0.17)$,
determined by the size-scaling.

\subsection{Open boundary condition for the insulator-insulator transition}

Next we consider the case of absent Rashba SOC: $\alpha=0$, i.e.,
the system belongs to the unitary class.
In this case, only direct transitions
between the two distinct topological insulators
are expected;
no metallic phase appears in between.
This is problematic to our numerical analyses, 
since fitting the data to identify a critical point is much more difficult
between two insulating phases.
Imagine that on both sides of the scale invariant critical point
$\lambda_L /L$ is a decreasing function of the size
as far as the system bears no surface states. 
To overcome this difficulty, we adopted, here in FIG. \ref{ll1},
{\it an open boundary condition},
since then the system can support a pair of gapless edge states
 in the quantum spin Hall phase, i.e., it becomes a metal-insulator transition.
Fitting data to determine a critical point is much easier in the vicinity of
a metal-insulator transition
than in insulator-insulator transitions.
\cite{ohtsuki99,obuse10}
The left panel of FIG. \ref{ll1} demonstrates 
determination of such a transition from a one-dimensional metal to 
an ordinary insulator ($E=0$, $\Delta=0$).

One might notice, however, presence of a stronger finite-size effect here
compared to the case of $\alpha \neq 0$ (FIG. \ref{ll2}, left panel).
Here, the existence of a scale invariant point is no longer evident,
and besides, apparent location of the focal point is much deviated from
that of the critical point: $(W_{\rm c}, \Lambda_0) = (7.461 \pm 0.046, 2.137 \pm 0.053)$ 
(indicated by dashed lines)
determined by the single-parameter scaling (fitting shown in the right panel).
Existence of such a stronger finite-size effect is a disadvantage
of 
an open boundary condition. 
\cite{ohtsuki99,obuse10}

\subsection{Open boundary condition for the weakly disordered regime}

Use of an open boundary condition has another area of utility.
In the weak disorder region,
the localization length $\lambda_L$ becomes large.
The larger $\lambda_L$ becomes, the larger system
one needs to consider to perform estimates of the localization length.
Use of an open boundary condition is useful for such a case,
since the typical length scale of the system,
i.e., the critical value of the localization length $\Lambda_0$ is 
smaller in system with an open boundary.
This is, on the other hand, applicable only for a metal to an {\it ordinary} 
insulator transition; e.g.,
in FIG. \ref{pd_e00}
only the phase boundary between OI and metal could be determined in this way.
This is because a metal-TI transition in an open boundary condition becomes metal-metal transition, which is also difficult to determine the critical point as insulator-insulator transition.

\section{Symmetry of the phase diagram}
\label{phs}

Let us finally comment on
the symmetry of the phase diagrams shown in Figs. \ref{pd_Delta-01}--\ref{pd_e05} with respect to $\Delta=4B$.
%
%
Let us define the local gauge transformation as 
\begin{align}
\hat \Xi c_{I,J} \hat \Xi^\dag = (-1)^{I+J} \sigma_x s_y c_{I,J},
\end{align}
which corresponds to $(\pi,\pi)$-shift in the momentum space, where $\bm r = (I,J)$ is a position on a square lattice.
With replacing $\Delta-4B \to -(\Delta-4B)$,  Hamiltonian Eq. (\ref{hmlt}) has symmetry as 
\begin{align}
\hat \Xi H|_{\Delta-4B \to -(\Delta-4B)} 
\hat \Xi^\dag = H.
\end{align}
Note that although disorder potential Eq. (\ref{dis}) is not microscopically invariant for the above transformation, disorder-averaged quantities such as a Green's function become invariant.
Consequently, the phase diagram becomes symmetric with respect to $\Delta=4B$.

\bibliography{phase_diagram}

\clearpage

\end{document}